\documentclass[12pt]{elsarticle}
\usepackage{graphicx}
\usepackage{bm}
\begin{document}
\title{Dynamical mass generation for ferromagnetic skyrmions in two dimensions}
\author{D. Wang}
\address{School of Science and Engineering, The Chinese University of Hong Kong, Shenzhen, Guangdong 518172, P. R. China}
\ead{dwwang@nudt.edu.cn}
\author{Hans-Benjamin Braun}
\address{Theoretical Physics, ETH Z\"{u}rich, CH-8093 Z\"{u}rich, Switzerland}
\author{Yan Zhou}
\address{School of Science and Engineering, The Chinese University of Hong Kong, Shenzhen, Guangdong 518172, P. R. China}
\ead{zhouyan@cuhk.edu.cn}
\begin{abstract}
Magnetic skyrmions are topological magnetization textures that are characterized by the homotopy group of two dimensional spheres. Despite years of intensive research on skyrmions, the fundamental problem of the inertia of a skyrmion in driven motion remains unresolved. By properly taking into account a direct coupling between skyrmion motion and the correspondingly excited magnons, we identify a dynamical mass for the skyrmion in motion. The direct coupling between skyrmion motion and magnons can be employed to engineer skyrmion dynamics with magnons through ingenious material and geometry design.
\end{abstract}
\maketitle
\section{Introduction}
\label{intro}
Magnetic skyrmions are magnetic topological solitons, the magnetization texture inside which induces a homeomorphic mapping between two dimensional (2D) spheres, defined as the 2D surface of solid spheres in three dimensions. The topological feature of magnetic skyrmions derives from the homeomorphic mapping and they are characterized by the homotopy group of 2D spheres \cite{Braun}, with the topological charge of an individual skyrmion defined as
\begin{equation}
Q = - \frac {1} {4 \pi} \int d^2 x \textbf{m} \cdot (\partial _x \textbf{m} \times \partial _y \textbf{m}).
\end{equation}
The spatial derivatives with respect to $x$ and $y$ are abbreviated as $\partial _x$ and $\partial _y$. $\textbf{m} = \textbf{M}/M _s$ is the unit magnetization vector, with $M _s$ the saturation magnetization and $\textbf{M}$ the magnetization vector distributed over a 2D plane. The concept of skyrmions was first proposed by Skyrme as a field theoretic description of hadrons \cite{Skyrme62}. In nanomagnetism, the existence of magnetic skyrmions was first postulated by Bogdanov and colleagues \cite{Bogdanov89, Bogdanov94, Rossler06} and later experimentally verified \cite{Muhlbauer09, Yu10}.

Similar to magnetic domain walls (DWs), magnetic skyrmions can serve as information storage elements and perform logic operations, due to their topologically protected stability against thermal agitations \cite{Fert13}. Additional advantages are derived from the lower threshold value in current density for current driven motion \cite{Jonietz10} and small sizes down to the atomic scale \cite{Heinze11}. As magnetic skyrmions in memory and logic devices need to be set to move to fulfill their functionalities, the skyrmion dynamics play a pivotal role in the understanding and application of these devices.

An important issue in determining skyrmion dynamics is the inertia, or mass, of an individual skyrmion. Time-resolved x-ray holography experiment \cite{Buttner15} measured a very large mass for the GHz dynamics of skyrmions, and inertial effects were also observed in micromagnetically simulated skyrmion diffusion \cite{Schutte14-2} and skyrmion dynamics driven by field gradients \cite{Moutafis09}. Theoretically, spontaneous symmetry breaking for the skyrmion position indicates that skyrmions should be massless \cite{Komineas15,Psaroudaki17}, if low temperature quantum effects \cite{Psaroudaki17} can be neglected. To reconcile the discrepancy between theory, experiment and simulation, Makhfudz \textit{et al.} \cite{Makhfudz12} phenomenologically derived a D\"{o}ring mass \cite{Doring} for skyrmionic bubbles, by considering an \textit{ad hoc}, long-range dipolar interaction for the bubbles, similar to the mass generation mechanism for a magnetic vortex confined in a nanodisk \cite{Guslienko10}. The effective dipolar interaction was introduced to simulate the energy cost induced by the deformation of skyrmions in motion. In principle, coupling to any other forms of deformation will give rise to a skyrmion mass, similar to the canonical Higgs mechanism for the mass generation for vector gauge bosons \cite{Higgs64,Englert64,Guralnik64}. Following this line, by resorting to spin-phonon interaction, Capic \textit{et al.} \cite{Capic20} derived a skyrmion mass that depends on the square of the magneo-elastic coupling constant. However, the most important, intrinsic interaction between skyrmions and magnons was investigated by Kravchuk \textit{et al.} \cite{Kravchuk18}, only to find a zero mass due to the absence of direct coupling between skyrmions and magnons.

Deploying a field theoretical method \cite{lag} to the same problem, we will identify a direct interaction between skyrmions and magnons, which is never discussed before and can induce a large dynamical mass for skyrmions in driven motion when the skyrmion size is large. In our treatment of the interacting skyrmion-magnon system, the skyrmion profile is determined by local minimization of the magnetic energy and is not deformed in the whole process of skyrmion motion. Any deviations from the metastable skyrmion profile are delegated into magnon modes. The magnons with non-zero energy can then be integrated out, bringing about inertial low energy dynamics for skyrmions. The dynamical mass thus obtained is a universal mass for skyrmion motion in the sense that the mass is an intrinsic property of skyrmions: the skyrmion mass is completely determined by the ground state profile of skyrmions and has nothing to do with the external stimuli used to drive skyrmions into motion; different stimuli excite different magnon modes and different skyrmion trajectories. An intrinsic mass, independent of the specific form of the external stimuli, is consistent with the picture of treating skyrmions as point particles. Therefore, our theoretical method has the advantage of being able to consistently and systematically consider the interaction between skyrmions and magnons, and can be easily generalized to discuss the motion of other topological magnetization textures.

The paper is organized as follows. In Sec. \ref{theo}, the Lagrangian description of magnetization dynamics is introduced. Then we expand the Lagrangian to solve for static skyrmion profile and magnon excitation spectrum in Secs. \ref{skyrprof} and \ref{magspectr} respectively. With the information for the skyrmion configuration and magnon excitation ready, we consider the coupling between skyrmoin motion and magnons in Sec. \ref{skyrmass} to derive a dynamical mass for skyrmion motion. The skyrmion dynamics under the presence of both damping and non-adiabatic spin transfer torque (STT) are treated in Sec. \ref{ab}. Sec. \ref{higgs} discusses the relation between our dynamical mass to the Higgs mechanism. Our conclusion is given in Sec. \ref{concl}. Detailed derivation of the direct coupling, explicit form for the magnon effective potential and the full action including dissipation after carrying out the path integral are given in \ref{surface}, \ref{potential} and \ref{pathint} respectively.
\section{Theory}
\label{theo}
The magnetization dynamics in the presence of both damping \cite{Gilbert04} and STT \cite{STT} are described phenomenologically by the Landau-Lifshitz-Gilbert (LLG) equation \cite{Landau80},
\begin{equation}
\partial _0 \textbf{m} = - \textbf{m} \times \textbf{h} + \alpha \textbf{m} \times \partial _0 \textbf{m} + \textbf{u} \cdot \nabla \textbf{m} - \beta \textbf{m} \times (\textbf{u} \cdot \nabla \textbf{m}),
\end{equation}
where we abbreviated the time derivative $\partial \textbf{m}/\partial t$ by $\partial _0 \textbf{m}$. The STT is characterized by an equivalent velocity $\textbf{u}$ \cite{Thiaville04} and a nonadiabatic parameter $\beta$ \cite{STT1}. As already stated in Sec. \ref{intro}, our calculation of skyrmion mass does not need the consideration of the external stimulus, here STT. But a previous treatment on the skyrmion motion driven by STT demonstrated no inertial behaviour \cite{Lin17}. We included the STT terms just to show why it is so, while skyrmions do have mass. The effective field $\textbf{h}$ is given by the functional derivative of the magnetic potential energy normalized to $2 A \lambda$, $W = \int d ^3 x w(\textbf{m}, \nabla \textbf{m})$, through $\textbf{h} = - \delta W /\delta \textbf{m}$.
\begin{equation}
w = \frac {(\partial _x \textbf{m}) ^2 + (\partial _y \textbf{m}) ^2} {2} + \frac {h_ D} {2} (m_z \nabla \cdot \textbf{m} - \textbf{m} \cdot \nabla m_z) - \frac {m_z^2} {2} - h _0 m _z
\end{equation}
is the magnetic potential energy density normalized to $2 |K|$, including the exchange, Dzyaloshinskii-Moriya (DM) \cite{DM}, magnetic anisotropy, and Zeeman interactions. $h_0$ is the external $z$ field normalized to the anisotropy field $H _K = 2 |K| / M _s$, and $h_D = D / \sqrt {A |K|}$ is the dimensionless DM field. The form of the DM energy density adopted here favors a N\'{e}el skyrmion configuration, with magnetization distributed over the $xy$ plane. In accordance with the dimensionless fields, the length and time are measured in terms of the DW width $\lambda = \sqrt{A/|K|}$ and the ferromagnetic resonance frequency $\omega _K = \gamma H _K$. $A$, $D$ and $K$ are exchange, DM and uniaxial anisotropy constants, respectively. We consider thin film systems with perpendicular magnetic anisotropy, so $K <0$. To take into account of the dipolar interaction, we can use the Green's function approach \cite{Guslienko00}. In the 2D case considered here, the dipolar interaction can be described by a local magnetostatic energy \cite{Guslienko11} and be absorbed into the magnetic anisotropy energy, resulting in an effective uniaxial anisotropy constant $K$.

The LLG equation is proven successful in description of macroscopic magnetization phenomena. However, to properly describe the interacting subsystems of magnetic solitons and the corresponding elementary excitations within the framework of micromagnetics, it is convenient to use the Lagrangian formulation of the LLG equation with the Lagrangian density \cite{Ivanov04,Guslienko10}
\begin{equation}
{\cal L} (\textbf{m}, \partial_\mu \textbf{m}) = (\textbf{n} \times \textbf{m}) \cdot \frac {\partial _0 \textbf{m} - \textbf{u} \cdot \nabla \textbf{m}} {1 + \textbf{n} \cdot \textbf{m}} - w(\textbf{m}, \partial_i \textbf{m}),
\end{equation}
complemented with the Rayleigh dissipation functional density
\begin{equation}
{\cal R} _d = \frac {\alpha} {2} (\partial _0 \textbf{m} )^2 - \beta (\textbf{u} \cdot \nabla \textbf{m}) \cdot \partial _0 \textbf{m},
\end{equation}
through the Euler-Lagrange equation
\begin{equation}
\partial _ 0 \frac {\delta L} {\delta (\partial _0 m _i)} - \frac {\delta L} {\delta m _i} + \frac {\delta R _d} {\delta (\partial _0 m _i)} = 0.
\end{equation}
In the Euler-Lagrange equation, Roman letters $i$ = 1, 2, 3 denote $x$, $y$, $z$ components respectively. We also used in ${\cal L}$ the convention that Greek letters denote numbers taking the values 0, 1, 2, and 3, with number 0 referring to the time component. Vector $\textbf{n}$ is a unit Dirac string vector \cite{Ivanov04}. For the treatment of a skyrmion with topological charge $Q = 1$, we can choose it to be along the $z$ direction, $\textbf{n} = \hat {z}$, which is also the direction of the magnetization vector at infinity.

For a static soliton, the magnetization profile of which is characterized by the polar angle $\theta$ and azimuthal angle $\phi$, a proper rotation of coordinate in the magnetization space can transform the magnetization vector into the direction of the local third axis. Under the rotation, the Lagrangian density becomes
\begin{equation}
{\cal L} (m _i, \partial_\mu m _i) = \frac {L _m ^{ln} n _l m _n} {1 + n _l m _l} (D _0 - u _i D _i) m _m  - w(m _k, D _i m _k),
\end{equation}
where the magnetization vector $\textbf{m}$ and the Dirac string vector $\textbf{n}$ are the corresponding vectors in the local rotated coordinate frames. As we will only work in the rotated coordinate frames, we do not differentiate between physical quantities in the original and rotated coordinate frames and use the same symbols for them. From now on, we stick to the Einstein convention, i.e. repeated indices are summed. $D _\mu = \partial _\mu - i A_\mu^i J_i$ is a covariant derivative for the magnetization vector \cite{Dugaev05}. Due to the rotation, there appears an emergent gauge field $A _\mu ^i$ acting on the rotated three dimensional magnetization field $\textbf{m}$. The Hermitian matrices $J _i$ are the generators for the SO(3) group and constructed from the fully antisymmetric Levi-Civita symbol, $i J_i^{jk} = \epsilon_{ijk}$.

To consider the interacting dynamics of the underlying soliton (skyrmion) and the corresponding elementary excitation (magnon) over the soliton profile separately, we decompose the magnetization into two parts, $m _i = \delta _3 ^i + s _i$, where the constant component along the third axis describes the soliton profile and the magnon excitation over the soliton is represented by small deviation amplitudes $s _i$. Using this decomposition, the kinetic part of the Lagrangian density can be expanded in transverse amplitudes $s _1$ and $s _2$ up to second order to give
\begin{equation}
{\cal L} _B = (1 - n _3) \tilde {\partial} _0 \phi - s _2 \tilde {\partial} _0 s _1 - \tilde {A} ^i s _i - \tilde {\partial} _0 \left(\frac {n _1 + n_3 s _1} {1 + n_3} s _2\right).
\label{lagb}
\end{equation}
$\tilde {\partial} _0 = \partial _0 - u _i \partial _i$ is the derivative in the frame of reference moving with the velocity $- \textbf{u}$, and the tilde gauge field is defined as $\tilde {A} ^i = A _0 ^i - u _j A _j ^i$. In ordinary formulation of Lagrangian density, total derivatives in time and space can be safely neglected since they only contribute to surface terms. However, in our assumption of magnons floating on top of a rigid soliton, the total time derivative can induce a contact interaction between magnons and the soliton's motion and should not be omitted, simply because the soliton center is time dependent and the differentiation on time can be transferred to a differentiation on the soliton's spatial coordinates. Then the spatial integration on the Lagrangian density to get the Lagrangian results in a surface term that couples the skyrmion motion and the magnon excitation (see \ref{surface}), thanks to the structure of the Dirac string potential, which diverges at the skyrmion center. Physically, this intrinsic coupling between skyrmion motion and magnons can be viewed as the emission of magnons by a moving skyrmion, with a Dirac string attached to its center. A similar picture has been discussed in the context of $\Delta \rightarrow \pi N$ decay \cite{Verschelde} and orbital dynamics of $^3$He-A \cite{Volovik}.
\section{Static skyrmion profile}
\label{skyrprof}
\begin{figure}
\begin{minipage}[c]{0.45\linewidth}\centering
\includegraphics[width=\linewidth]{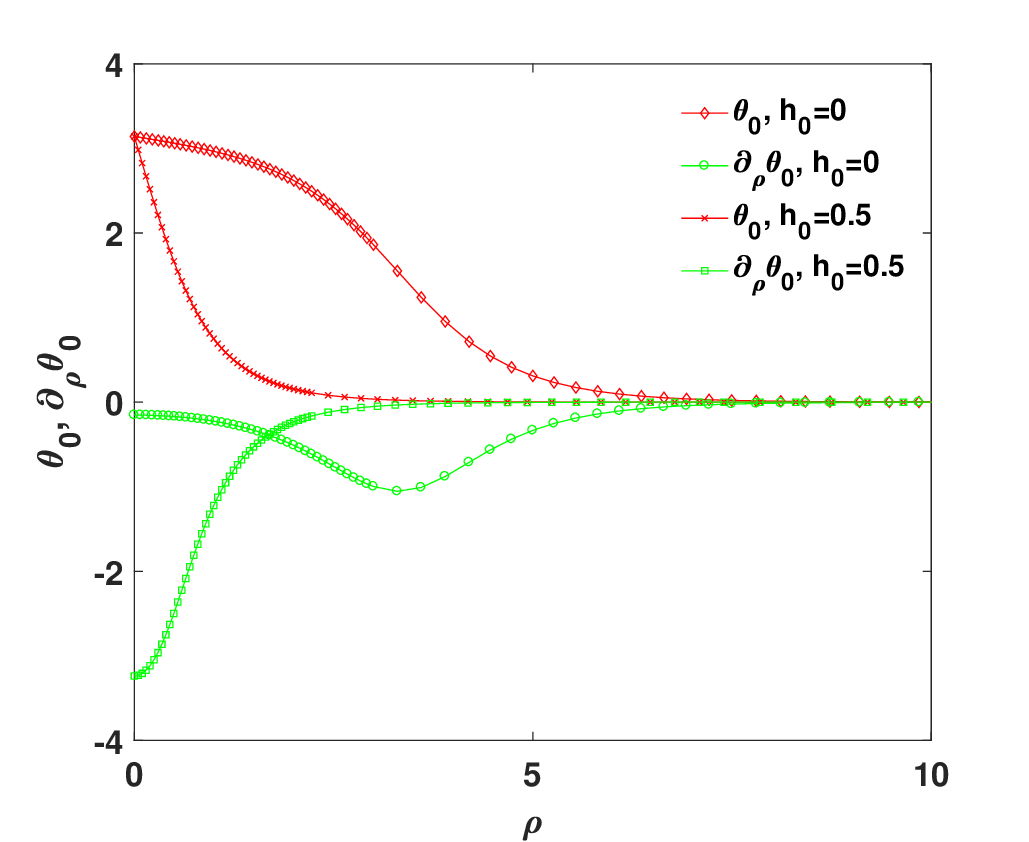}
\end{minipage}
\caption{Skyrmion profile $\theta_0$ and its derivative on radial coordinate $\rho$, $\theta'_0$, for $h _D = 1.2$, with two values of the applied external field shown in the legend.}
\label{profile}
\end{figure}

To proceed further, we need to designate the skyrmion profile and then to determine the corresponding magnon spectrum. The skyrmion profile $\theta_0$ is determined by the zeroth order energy density
\begin{equation}
w_0 = \frac {{\theta' _0} ^2} {2} + \frac {n _1 ^2} {2 \rho ^2} + \chi \frac {h _D} {2} \left(\theta' _0 - \frac {n _1 n _3} {\rho}\right) - \frac {n _3 ^2} {2} - n _3 h _0
\end{equation}
using cylindrical coordinates $\rho$ and $\varphi$ centered around the skyrmion center. $n _1 = - \sin \theta _0$, $n _3 = \cos \theta _0$, and a prime denotes the radial derivative, $\theta' _0 = d \theta _0/d \rho$. In deriving $w_0$, we assumed that the skyrmion has rotational symmetry around its center and $\phi _0 = \varphi + \varphi_0$, with $\varphi_0$ = 0 or $\pi$ defining the chirality $\chi = \cos \varphi _0$ of the skyrmion. The skyrmion profile is determined by a variation of $w _0$ with respect to the angle $\theta _0$,
\begin{equation}
\theta'' _0 + \frac {\theta' _0} {\rho} + n _1 n _3 \frac {\rho^ 2 + 1} {\rho^ 2} + \chi h_D \frac {n _1 ^2} {\rho} + h_0 n _1 = 0.
\end{equation}
Representative skyrmion profiles obtained using a shooting method \cite{Bogdanov99} are shown in Fig. \ref{profile}.
\section{Magnon excitation spectrum}
\label{magspectr}
Introducing the real spinor wave function $\psi ^T = (s _1, s _2)$, the second order Lagrangian density becomes
\begin{equation}
{\cal L} _2 = \frac {i} {2} \psi ^\dagger \sigma _2 \partial _0 {\psi} + \psi ^ \dagger \frac {\textbf{D} ^2} {2} \psi - \psi ^\dagger \left(\frac {v _0} {2} + \frac {v _3} {2} \sigma _3\right) \psi
\end{equation}
where $v_0$ and $v _3$ are effective potentials \cite{Schutte14-1} (see \ref{potential}). $\sigma _2$ and $\sigma _3$ are the second and third Pauli matrices. The magnon eigenmode is determined by the application of the Euler-Lagrange equation to ${\cal L} _2$, leading to
\begin{equation}
i \sigma_2 \partial _0 {\psi} = (v_0 + v_3 \sigma _3 - \textbf{D} ^2) \psi.
\end{equation}
We assume that the magnon excitation has the form \cite{Kravchuk18} $s_1 = f _m (\rho) \cos (m \varphi + \omega t)$ and $s_2 = g _m (\rho) \sin (m \varphi + \omega t)$. Then the equation of motion for the magnon amplitude $\psi _m ^T = (f _m, g _m)$ becomes $\omega \sigma _1 \psi _m = H_m \psi _m$, where the Hamiltonian is $H _m = v_0 + v_3 \sigma_3 - \textbf{D} _m ^2$. The covariant derivative is
\begin{equation}
\textbf{D} _m = \nabla _m - \sigma _1 \left(\frac {n _3} {\rho} + \chi \frac {h _D} {2} n _1 \right) \hat {\varphi}
\end{equation}
in cylindrical coordinates fixed to the skyrmion center. $\sigma _1$ is the first Pauli matrix, and $\nabla _m$ is obtained by substituting $\partial _\varphi$ with $m$ in $\nabla$. Due to the presence of the emergent SU(2) gauge field for magnons \cite{Dugaev05}, there exists a Berry phase associated with the emergent gauge field for magnons circulating around a skyrmion.

The Hamiltonian $H_m$ is invariant under the joint transformation $m \rightarrow -m$ and a rotation by $\sigma_3$ in the spinor space \cite{Psaroudaki17}, $H_m = \sigma _3 H_{-m} \sigma _3$. Correspondingly, the magnon eigenequation is invariant under the joint transformation $\omega \rightarrow -\omega$, $m \rightarrow -m$, and $\psi _m \rightarrow \sigma _3 \psi _ {-m}$. This transformation corresponds to the particle-hole symmetry for the original wave function, $\psi \rightarrow \sigma_3 \psi ^*$, which is reminiscent of the time reversal symmetry of the LLG equation without damping and STT. A similar particle-hole symmetry was also found for electrons \cite{Wang20} and magnons \cite{Wang17} in magnetic DWs. Due to the presence of the particle-hole symmetry, the solutions for the eigenequation with positive and negative frequency are related to each other, through relation $\psi _ m (\omega) \propto \sigma_3 \psi _{-m} (- \omega)$. Thus we can retain only the positive frequency spectrum \cite{Kravchuk18}.

\begin{figure}
\begin{minipage}[c]{0.45\linewidth}\centering
\includegraphics[width=\linewidth]{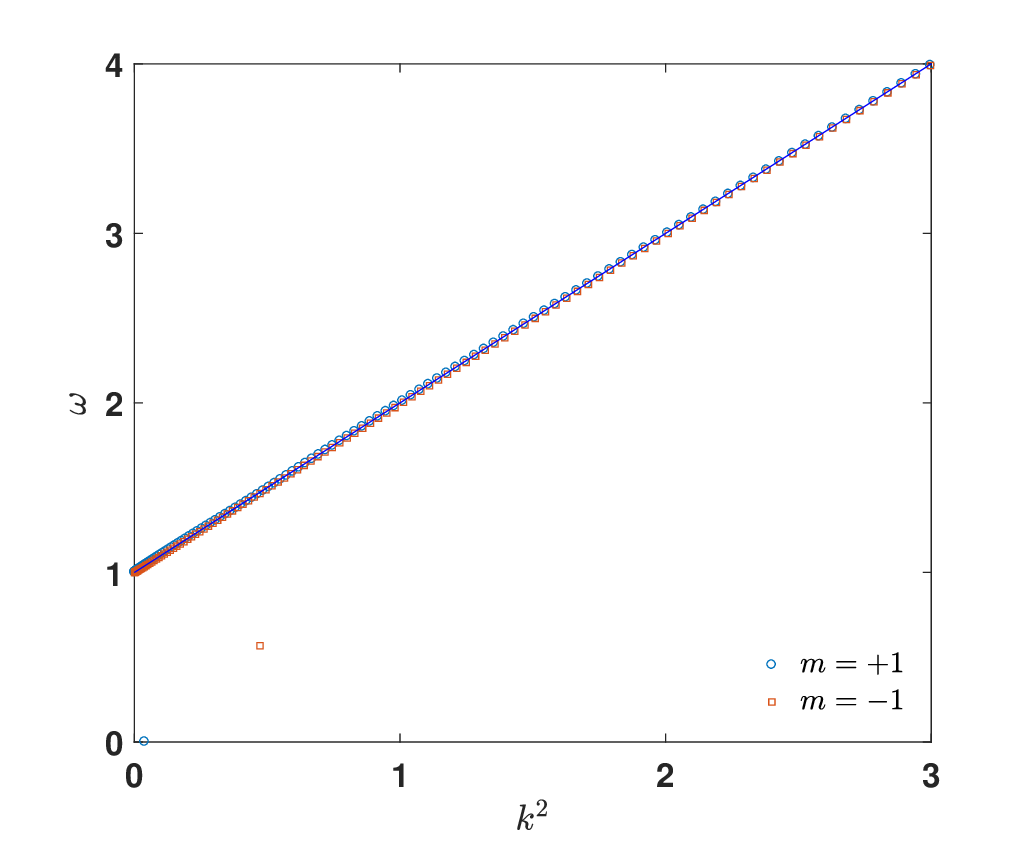}
\end{minipage}
\caption{Zero field magnon dispersion with respect to the effective radial wave number $k$, which is calculated as the average value, with weight function $\sigma _1$, of the wave number appearing in the Bessel function expansion for $\psi _m$. The zero energy state in the $m=1$ spectrum corresponds to the Goldstone mode, while the state in the continuum gap of the $m=-1$ branch is a localized state, which will disappear as the skyrmion is shrunk by an applied field. The solid line is $1 + k ^2$.}
\label{dispersion}
\end{figure}

The magnon spectrum can be obtained by expanding $\psi _m$ in terms of Bessel functions \cite{Schutte14-1,Lin17}. As shown in Fig. \ref{dispersion}, the obtained spectrum can be labeled by the frequency eigenvalue $\omega _m ^i$, with corresponding wave function $\psi _m ^i$. Then, similar to the eigenvalue problem of the Schr\"{o}dinger equation, we can use the eigenequation for the magnon excitation to discuss the orthogonality of $\psi _m$. The obtained orthonormal relation for the real amplitude wave function is
\begin{equation}
\int \rho d \rho (\psi _m ^i) ^\dagger \sigma _1 \psi_m ^j = 4 \delta _i ^j;
\end{equation}
eigenfunctions with different eigenvalues are orthogonal to each other with respect to the weight function $\sigma_1$. We chose to normalize the wave functions to 4, instead of the usual unity, in order to be consistent with the normalization of the Goldstone mode \cite{Goldstone}, which appears as a zero frequency excitation in the $m$ = 1 spectrum.
\section{Effective mass for skyrmions}
\label{skyrmass}
With the obtained skyrmion profile and magnon spectrum, we are now ready to investigate the dynamics of the interacting skyrmion-magnon system. For this purpose, we assume that the skyrmion is rigid during its motion, with the moving profile characterized by $\theta = \theta _0 (\bm{\rho} - \bm{\rho} _c)$ and $\phi = \phi _0 (\bm{\rho} - \bm{\rho} _c)$, and the motion of the rigid skyrmion is just a displacement of the skyrmion center $\bm{\rho} _c$. We then decompose the magnon excitation superimposed on top of the moving skyrmion into a set of dynamical variables through $s_1 = f_m ^i (\rho) [a _m ^i (t) \cos m \varphi + b _m ^i (t) \sin m \varphi]$ and $s_2 =  g_m ^i (\rho) [a _m ^i (t) \sin m \varphi - b _m ^i (t) \cos m \varphi]$. In terms of this expansion, the radial function of the Goldstone mode has the form $f _1 ^0 = - \theta' _0$ and $\rho g _1 ^0 = \sin \theta _0$ \cite{Kravchuk18}. The spatial variables $\rho$ and $\varphi$ appearing in the expansion are measured with respect to the center of the moving skyrmion. The superscript $i = 0$ refers to the Goldstone or localized states in the magnon spectrum, so the states in the continuum spectrum start with superscript $i = 1$, as there is at most one localized state in the magnonic gap \cite{Kravchuk18}.

Substitute in the expansion for $s _i$ and perform the spatial integration, we can get the Lagrangian normalized to $2 \pi \tau$, where $\tau$ is the thickness of the magnetic film,
\begin{equation}
\mbox {\hspace{-1cm}} L = (\dot {p} _1 ^0 + v _c) ^\dagger i \sigma _2 p _1 ^0 + (\dot {p} _m ^i) ^\dagger i \sigma _2 p _m ^i - \omega _m ^i (p _m ^i) ^\dagger p _m ^i + v _c ^\dagger (\eta _i i \sigma _2 {p} _1 ^i + \bar {\eta} _i \sigma _1 p _ {-1} ^i)
\label{lag}
\end{equation}
in terms of the spinor magnon amplitudes $p _m ^i = (a _m ^i, b _m ^i) ^T$. $v _c = \dot {\rho} _c + u$ is the velocity of the skyrmion center in the reference frame moving with velocity $-u$ where $ u ^T= (u _1, u _2)$ and $\dot {\rho} _c ^T = (\dot {\rho} _c ^1, \dot {\rho} _c ^2)$. A dot over symbols is used to denote the time derivative, $\dot {\rho} _c = d \rho _c/dt$, and the summation excludes the Goldstone mode. Constants $\eta _i = g _1 ^i (0)/ \theta' _0(0)$ and $\bar {\eta} _i = g _ {-1} ^i (0)/ \theta'_0(0)$ characterize the coupling between the skyrmion motion and the magnon excitation, which depends only on the values of the functions $g _{\pm 1} ^i$ and $\theta' _0$ at the skyrmion center, as discussed in Sec. \ref{theo}. The sum in (\ref{lag}) includes gapped magnon states only, excluding the Goldstone mode, the dynamics of which are explicitly given in the first term. The second and third terms describe the dynamics of the magnon subsystem, while the interplay between the skyrmion motion, STT and magnons arises from the fourth term. Due to the coupling between the skyrmion motion and the sea of $|m| = 1$ magnons, the would-be instantaneous response of a skyrmion to the sudden switching-on of STT is impeded by its coupling to the magnon degree of freedom. Correspondingly, the motion of the skyrmion will inevitably excite magnons and the skyrmion motion is accompanied by emission of magnons.
\begin{figure}
\begin{minipage}[c]{0.45\linewidth}\centering
\includegraphics[width=\linewidth]{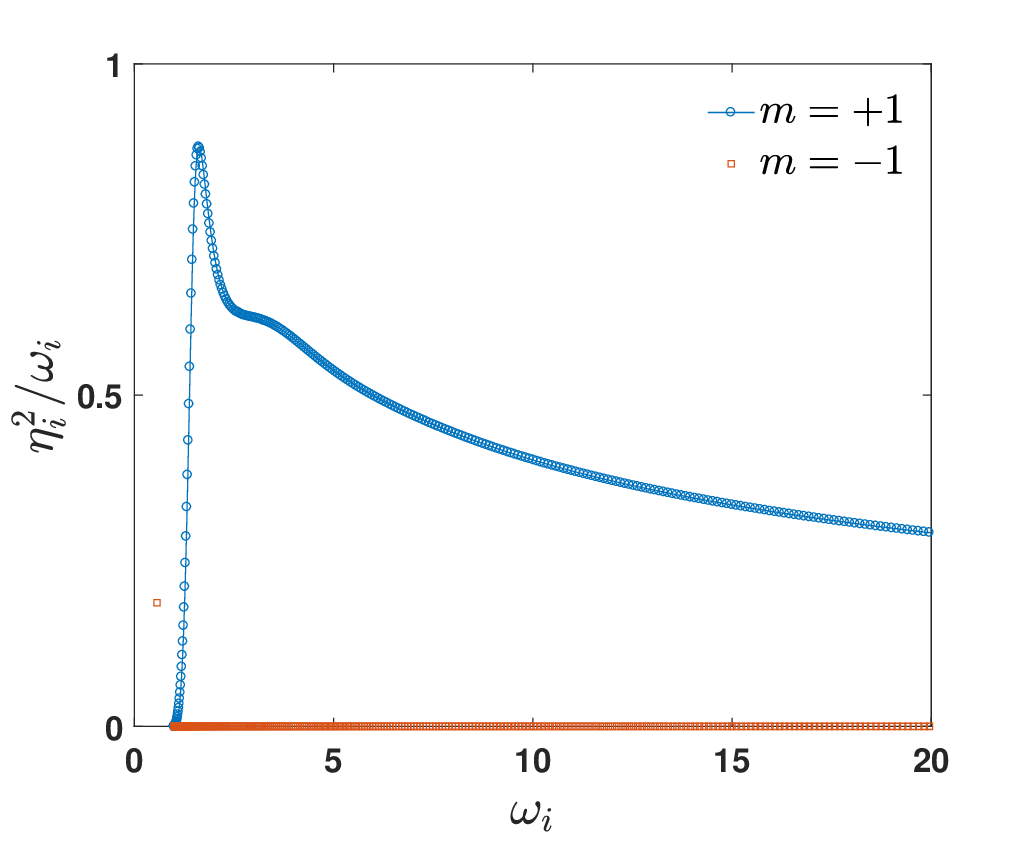}
\end{minipage}
\caption{Mass spectrum for $h _D = 1.2$ with zero applied field. For the $m = -1$ magnon branch, only the localized state inside the magnon gap contributes significantly to the skyrmion mass; the continuum's contribution is negligible.}
\label{mspectrum}
\end{figure}

Due to the coupling between the skyrmion motion and the $|m| = 1$ magnons, the motion of a skyrmion is not massless anymore. This fact can be seen easily if we describe the dynamics of the skyrmion-magnon interacting system in the form of a path integral. In the path integral formulation of the dynamics \cite{Altland,Kamenev} using magnon coherent states, the dynamical variables can be taken to be the magnon amplitudes $a _m ^i$ and $b _m ^i$, instead of the commonly adopted spatial variables. Then, any path in the phase space connecting the initial and final states will contribute a phase factor $\exp (i S/ \hbar)$ to the path integral, or propagator, where $S = \int dt L$ is the action. The phase factor measures the relative importance of the path considered in determination of the actual path, as it evolves in the phase space. The magnon degree of freedom in the path integral formulation can be integrated out (See \ref{pathint}) to achieve an effective Lagrangian (in units of $2 \pi \tau$) for the skyrmion motion
\begin{equation}
L _{eff} = (\dot {p} _1 ^0 + v _c) ^\dagger i \sigma _2 p _1 ^0 +  \frac {\mu} {2} v _c ^\dagger v _c,
\label{lageff}
\end{equation}
neglecting surface and correlation contributions to the action. As the Goldstone mode $p _1 ^0$ has zero energy, it cannot be integrated out and is kept in (\ref{lageff}).

The effective Lagrangian (\ref{lageff}) shows that the skyrmion becomes massive. An identical result can be obtained by considering the classical equation of motion for the skyrmion motion, through eliminating the magnon amplitudes using their equations of motion. The resultant skyrmion mass $\mu$ is proportional to the square of the coupling constants divided by the magnon frequency,
\begin{equation}
\mu = \frac {\eta _i ^2} {2 \omega _1 ^i} + \frac {\bar {\eta} _i ^2} {2 \omega _{-1} ^i}.
\end{equation}
Since the coupling constants are given by the ratio of $g _ {\pm 1} ^i (0)$ to $\theta' _0 (0)$, the physical meaning of the skyrmion mass is obvious: $\theta _0 '(0)$ characterizes the intrinsic degree of deformation, while $g _ {\pm 1} ^i (0)$ gives the agitation to the static skyrmion profile; so the competition of the two tendencies, i.e. one resists and the other favors change of the skyrmion profile, determines the mass $\mu$. The calculated mass spectrum is shown in Fig. \ref{mspectrum}, where we can see that the skyrmion mass is mainly determined by the $m = 1$ magnon continuum and the $m = -1$ localized state, while the $m = -1$ continuum contribution is negligible.

\begin{figure}
\begin{minipage}[c]{0.45\linewidth}\centering
\includegraphics[width=\linewidth]{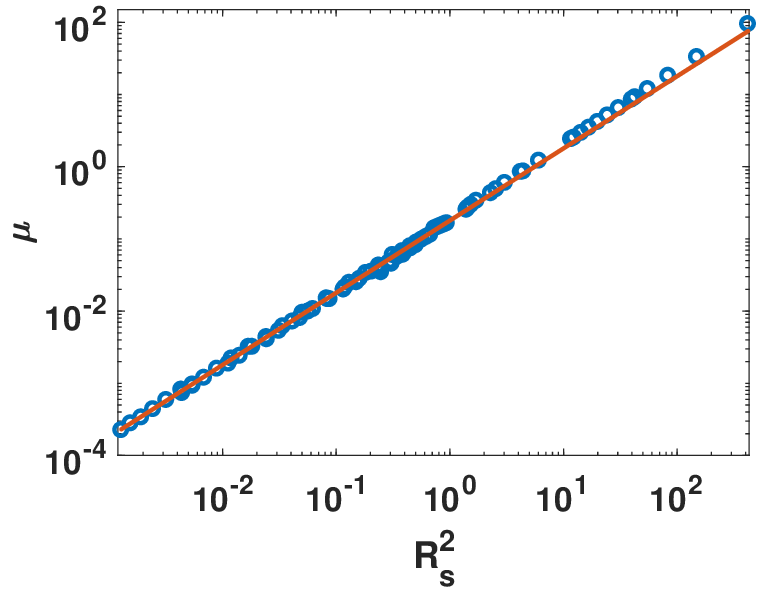}
\end{minipage}
\caption{Skyrmion mass $\mu$ as a function of the skyrmion radius squared, $R _s ^2$. The solid line is a linear fit to the mass data with zero interception. The skyrmion radius $R _s$ is controlled by a combination of the DM field $h _D$ and the external field $h _0$. It should be noted that the correspondence between $R _s$ and $h _D$, $h _0$ is not one to one; there are more than one combination of $h _D$ and $h _0$ corresponding to one $R _s$.}
\label{mass}
\end{figure}

Close inspection of the mass spectrum reveals a problem: As the frequency increases, the mass spectrum decays too slowly to give a finite total mass. The divergence of the total mass is logarithmic in frequency. The divergence arises because of our continuum description of magnetization dynamics, which is only valid for low frequency magnons. As the frequency is increased, the discrete nature of the underlying crystal lattice will become more important and the continuum description fails. A similar ultraviolet divergence was observed in micromagnetic analysis for the mass of magnetic vortices \cite{Wysin94}. To resolve this problem, we need to consider the magnon spectrum on discrete lattices, which is beyond the scope of the current work. We can circumvent this awkward situation by imposing a cut-off frequency $\omega _c$ for the frequency summation. By employing $\omega _c = 100$, the obtained total mass as a function of the square of an effective radius $R_s$ for the skyrmion, which is determined through $R _s = \pi/|\theta'_0 (0)|$, is displayed in Fig. \ref{mass} for $h _D$ ranging from 1.2 to 0.2 and various $h _0$. As can be seen, the skyrmion mass decreases rapidly with the decrease of $R_s$, and the skyrmion mass is almost linearly proportional to the skyrmion area $\pi R _s ^2$, which measures effectively how many spins are enclosed in the skyrmion. The actual unit for $\mu$ is $4 \pi A \tau/ \lambda ^2 \omega _K ^2$. For $A = 10$ pJ/m, $\lambda = 10$ nm and $\omega _K = 2 \pi$ GHz, the mass unit for a $\tau = 1$ nm thick film is $10 ^{-22}/ \pi$ kg. As $h _D = 1.2$ in Fig. \ref{mass} is very close to the critical value $4/\pi$ \cite{Rohart13}, the skyrmion size is rather large, giving rise to the large mass shown in Fig. \ref{mass} with $h _0 = 0$.

Skyrmion kinetic energy in $L _ {eff}$ is actually proportional to the square of $v _c$, indicating that skyrmion motion driven by STT costs no energy and appears massless \cite{Lin17} if $\dot {\rho} _c = - u$, even if the actual dynamical mass $\mu$ is not zero. However, we would like to emphasize that this apparent instantaneous response to adiabatic STT is only rendered possible by the presence of a dynamical mass for skyrmion motion, imposing an energy penalty for any lag in response. The physics for this counterintuitive phenomenon is simple: If the skyrmion moves together with the itinerant electrons, the electrons will see no variation of magnetization and hence no STT will arise; The adiabatic STT is simply transformed out to the motion of a reference frame fixed to the center of the skyrmion. This fact demonstrates that theoretical discussions based on the dynamical equations of motion for the skyrmion center \cite{Schutte14-2,Lin17} cannot be used to determine the skyrmion mass unequivocally, since an instantaneous response of a skyrmion does not guarantee that the skyrmion is massless, especially so in the case of skyrmion motion driven by STT. The instantaneous dynamics of skyrmion motion driven by adiabatic STT only will be changed to apparent massive dynamics in the presence of both damping and nonadiabatic STT for skyrmions with large $R_s$, and Goldstone and localized magnonic modes will be excited in the steady state, as will be shown in the following section, Sec. \ref{ab}.

\section{Dynamics with damping and nonadiabatic STT}
\label{ab}
To see how the instantaneous behaviour of skyrmion motion is modified by damping and nonadiabatic STT, we need the lowest order Rayleigh dissipation functional in units of $\pi \tau$
\begin{equation}
\mbox {\hspace{-1cm}} R _d = c _1 ^ 0 \left[\frac {\alpha} {4} \dot {\rho} _c ^\dagger \dot {\rho} _c + \alpha (\dot {p} _1 ^0) ^\dagger \dot {p} _1 ^0 + \frac {\beta} {2} u ^\dagger \dot {\rho} _c + v _d ^\dagger \dot {p} _1 ^0 \right] + \alpha c _n ^ i (\dot {p} _n ^i) ^\dagger \dot {p} _n ^i + \bar {c} _0 v _d ^\dagger \sigma _3 \dot {p} _{-1} ^0,
\label{rayleigh}
\end{equation}
where the coupling constants are defined as
\begin{equation}
c _n ^ i = \int \rho d \rho (\psi _ n ^i) ^ \dagger \psi _ n ^i
\end{equation}
and
\begin{equation}
\bar {c} _ 0 = \int \rho d \rho (\psi _ 1 ^0) ^ \dagger \sigma _3 \psi _ {-1} ^0.
\end{equation}
$v _d = \alpha \dot {\rho} _c + \beta u$ is a dissipation velocity. Negligible contributions to $R _d$ are discarded. Substitute the Lagrangian (\ref{lag}) and the Rayleigh dissipation functional (\ref{rayleigh}) into the Euler-Lagrange equation, we get the equation of motion for the skymion center as
\begin{equation}
\mu \dot {v} _c = \frac {i} {2} \sigma _2 v _c - i \nu \sigma _2 \dot {v} _d - f _i y ^i - \bar {f} _0 \sigma _3 \bar {y} ^0,
\end{equation}
with constants $\nu = \bar {c} _0 \bar {\eta} _0/ 4 \omega _{-1} ^0$, $ f _i = {\eta _i}/ {\omega _1 ^i}$, and $\bar {f} _0 = {\bar {\eta} _0}/ {\omega _{-1} ^0}$. As can be expected, the mass $\mu$ appears naturally. The reactive part of the equation of motion can be reproduced by an effective Lagrangian
\begin{equation}
L _{eff} = (\dot {p} _1 ^0 + v _c) ^\dagger i \sigma _2 p _1 ^0 + \frac {(\dot {q} _1 ^i) ^\dagger \dot {q} _1 ^i} {4 \omega _1 ^i} + \frac {(\dot {q} _{-1} ^i) ^\dagger \dot {q} _ {-1} ^i} {4 \omega _ {-1} ^i},
\end{equation}
where $\dot {q} _1 ^i = 2 \dot {p} _1 ^i + \eta _i v _c$, $\dot {q} _ {-1} ^i = 2 \dot {p} _ {-1} ^i + \sigma _3 \bar {\eta} _i v _c$, and the summation in $i$ excludes the Goldstone mode. The functions
\begin{equation}
y _i = \left(1 - \frac {\alpha} {2} c _1 ^ i i \sigma _2\right) \ddot {p} _1 ^i
\end{equation}
and
\begin{equation}
\bar {y} _0 = \left(1 - \frac {\alpha} {2} c _{-1} ^ 0 i \sigma _2 \right) \ddot {p} _{-1} ^0
\end{equation}
are related to the second order time derivative of the $|m| = 1$ magnon amplitudes. The process of eliminating the magnonic degree of freedom can be repeated to introduce terms that are higher than $\dot {v} _c$ in time derivative, through elimination of $\ddot {p} _1 ^i$ and $\ddot {p} _{-1} ^0$.

The corresponding transient dynamics of skyrmions are very complicated, involving the excitation of all the magnon eigenmodes. This agitation of the magnon sea will generally induce delay in the skyrmion's response to STT. The delay in skyrmion's response relies critically on the existence of localized states in the $m = - 1$ magnonic band gap. Only skyrmions with large size can host localized states, and the inertial effect of these large skyrmions is observable. If the skyrmion size is either shrunk by an external field or a small DM coupling constant, the response will still be instantaneous due to the absence of localized states in the magnonic band gap, although the dynamical mass is always there. The study of the transient dynamics further reveals that the main effect of the dynamical mass is to determine the low frequency pole, or the frequency of low-energy quasiparticles, of the interacting skyrmion and magnon system, while the existence of localized states in the magnonic band gap determines whether the response is instantaneous. An instantaneous response to STT does not guarantee zero mass.

Due to the presence of the Gilbert damping, the transient magnon excitation amplitudes will relax to constant values that can be easily obtained,
\begin{equation}
p _{-1} ^0 = \bar {c} _ 0 \frac {\alpha - \beta} {4 \omega _ {-1} ^0} \sigma _3 u
\end{equation}
and
\begin{equation}
\dot {p} _1 ^0 = c _1 ^0 \frac {\beta - \alpha} {4} i \sigma _2 u,
\end{equation}
accompanying the uniform motion of the skyrmion center $\dot {\rho} _c = - u$. The main effect of the translational motion of the skyrmion center is to excite the localized $m = -1$ mode and the Goldstone mode. The excitation amplitudes of all other modes are negligibly small, due to the orthogonality between the propagating magnon modes and the Goldstone mode. Irrespective of the restrictions imposed by the orthogonality, the excitation amplitudes are proportional to $\omega _i ^ {-1}$, so the excitation of high frequency magnons by skyrmion translation is dynamically suppressed.

If the skyrmion motion is characterized by the position vector averaged over the topological charge density \cite{Kravchuk18},
\begin{equation}
- \int d^2 x \frac {\textbf{m} \cdot (\partial _x \textbf{m} \times \partial _y \textbf{m})} {4 \pi Q} \bm{\rho},
\end{equation}
then the skyrmion velocity is given by the sum $\dot {\rho} _c + \dot {p} _1 ^0$, instead of $\dot {\rho} _c$. As $\dot {p} _1 ^0$ is perpendicular to $u$, the skyrmion motion is deflected from the motion of itinerant electrons. The deflection, dubbed the skyrmion Hall effect \cite{Zang11,Nagaosa13}, is purely mediated by the Goldstone mode. In the case of $\beta = \alpha$, the deflection is absent and the skyrmion profile travelling at velocity $-u$ is an exact solution of the LLG equation in the presence of both damping and STT.

\section{Relation to the Higgs mechanism}
\label{higgs}
Despite the superficial resemblance between the dynamical mass generation process presented here and the renowned Higgs mechanism responsible for the acquirement of mass for massless gauge bosons \cite{Higgs64,Englert64,Guralnik64}, they are actually not identical mechanisms for mass generation and should not be confused with each other. In the canonical Higgs mechanism, gauge bosons acquire a mass through interaction with Goldstone bosons. The Goldstone bosons disappear after a redefinition of the gauge bosons, similar to our redefinition of the skyrmion velocity, but the generated mass is proportional to the expectation value of the vacuum with spontaneous symmetry breaking. In our mechanism for dynamical generation of mass, the mass is just a consequence of the interaction between the skyrmion motion and the massive vacuum excitations. This mechanism is just an explicit demonstration of the equivalence between energy and mass. The physical picture behind this mechanism is simple: The motion of skyrmions will inevitably induce excitation of magnons, and it is the back action of the magnons that prevents skyrmions to respond instantaneously to the external stimuli, endowing inertia to skyrmions. Although the magnonic Goldstone modes can be absorbed into the skyrmion translation, the dynamical mass has nothing to do with the spontaneous symmetry breaking in translation. Nevertheless, it is interesting to note that the second model discussed by Englert and Brout \cite{Englert64} is more relevant to our discussion: The gauge bosons are the Goldstone bosons at the same time, and the mass derived from spontaneous symmetry breaking is proportional to the squared product of the coupling constant and the mass of the fermions. The only modification needed to reproduce the correct scaling behaviour of our dynamical $\mu$ is to use the magnon mass $\omega _i ^ {-1}$, instead of the mass squared.
\section{Conclusion}
\label{concl}
By investigating the direct interaction between skyrmions in motion and the magnons floating on top of them, we derived a universal dynamical mass for the driven motion of individual skyrmions. In the case of skyrmion motion driven by pure adiabatic STT, the dynamical mass manifests itself by requiring an instantaneous response to the STT, rendering the apparent skyrmion motion non-inertial due to the unique coupling mechanism between skyrmion motion and adiabatic STT. It has the usual inertial effect only in the skyrmion dynamics for skyrmions with large radius in the presence of both damping and nonadiabatic STT. Due to the same interaction that gives rise to the dynamical skyrmion mass, the motion of skyrmions will inevitably excite magnons, although effective only for Goldstone and localized states in the long time limit.
\section*{Acknowledgements}
D.W. would like to express gratitude for hospitality to Department of Applied Physics, University of Gothenburg where the current work was initiated. Y.Z. acknowledges the support by Guangdong Special Support Project (2019BT02X030), Shenzhen Peacock Group Plan (KQTD20180413181702403), Pearl River Recruitment Program of Talents (2017GC010293) and National Natural Science Foundation of China (11974298, 61961136006).
\appendix
\section{Direct coupling between skyrmion motion and magnons}
\label{surface}
For the discussion of the direct coupling between skyrmion motion and magnons, it suffices to consider just the $\partial _0$ component of the last term in the Lagrangian density (\ref{lagb}), as the term proportional to $u _i \partial _i$ leads to coupling to adiabatic STT only. According to our definition for the unit Dirac string vector in the local, rotated, coordinate frame, we have the explicit expressions $n _1 = - \sin \theta _0 (\rho)$ and $n _3 = \cos \theta _0 (\rho)$, so the Dirac string vector potential is
\begin{equation}
A _s (\rho) = \frac {n _1} {1 + n _3} = - \tan \frac {\theta _0 (\rho)} {2},
\end{equation}
where we have shown explicitly the dependence of the polar magnetization angle, $\theta _0$, on the radial distance, $\rho = \sqrt {x _r ^2 + y _r ^2}$, away from the skyrmion center specified by $\rho _c ^1 (t)$ and $\rho _c ^2 (t)$. $x _r = x - \rho _c ^1 (t)$ and $y _r = y - \rho _c ^2 (t)$ just measure the relative distance from the skyrmion center. To simplify the notation, we write the spin wave amplitude as $s_2 = g_m ^i (\rho) \psi _m ^i (\varphi, t)$, where $\psi _m ^i (\varphi, t) = a _m ^i (t) \sin m \varphi - b _m ^i (t) \cos m \varphi$ and summation over repeated indices are implied. Similar to the radial coordinate $\rho$, the polar angle $\varphi$ is also defined relative to the skyrmion center, $\varphi = \tan ^{-1} (y _r/ x _r)$.

With those definitions, the calculation of the Lagrangian density is straightforward,
\begin{eqnarray}
\partial _0 (A _s s _2) &=& A _s (\rho) g_m ^i (\rho) \partial _0 \psi _m ^i (\varphi, t) + \psi _m ^i (\varphi, t) \partial _0 \rho \partial _\rho [A _s (\rho) g_m ^i (\rho)]\nonumber\\
&+& A _s (\rho) g_m ^i (\rho) \partial _0 \varphi \partial _\varphi \psi _m ^i (\varphi, t).
\end{eqnarray}
The time derivative in the first term acts only on the dynamical variables $a _m ^i$ and $b _m ^i$. After the integration over space, it amounts to a total time derivative contribution to the Lagrangian, and thus can be neglected. The third term, which arises due to the time dependence of the skyrmion center, can be simplified by first considering the angular integration
\begin{eqnarray}
\int d \varphi \partial _0 \varphi \partial _\varphi \psi _m ^i (\varphi, t) = \int \frac {d \varphi} {\rho} \partial _\varphi \psi _m ^i (\varphi, t) \partial _\varphi (\partial _0 \rho) = \partial _0 \rho \left. \frac {\partial _\varphi \psi _m ^i (\varphi, t)} {\rho} \right| _0 ^ {2 \pi}\nonumber\\
- \int \frac {d \varphi} {\rho} \partial _0 \rho \partial _\varphi ^2 \psi _m ^i (\varphi, t)
= m ^2 \int \frac {d \varphi} {\rho} \psi _m ^i (\varphi, t) \partial _0 \rho = \int d \varphi \psi _m ^i (\varphi, t) \frac {\partial _0 \rho} {\rho} ,
\label{pint}
\end{eqnarray}
where the relation $\rho \partial _0 \varphi = \partial _\varphi (\partial _0 \rho)$, which can be easily verified by explicitly calculating the time derivatives
\begin{equation}
\partial _0 \rho = - \frac {x _r \dot {\rho} _c ^1 (t) + y _r \dot {\rho} _c ^2 (t)} {\rho} = - \dot {\rho} _c ^1 (t) \cos \varphi - \dot {\rho} _c ^2 (t) \sin \varphi
\end{equation}
and
\begin{equation}
\partial _0 \varphi = - \frac {x _r \dot {\rho} _c ^2 (t) - y _r \dot {\rho} _c ^1 (t)} {\rho ^2} = - \frac {\dot {\rho} _c ^2 (t) \cos \varphi - \dot {\rho} _c ^1 (t) \sin \varphi} {\rho} = \frac {\partial _\varphi (\partial _0 \rho)} {\rho},
\end{equation}
has been used in the first step. In the lase step, the fact that only $m = \pm 1$ magnon modes contribute to the integral has been invoked to set $m ^2 = 1$. Similarly, the second term can be simplified by considering the radial integration first
\begin{eqnarray}
\mbox {\hspace{-2cm}} \int \rho d \rho \partial _0 \rho \partial _\rho [A _s (\rho) g_m ^i (\rho)] = \left. \rho A _s (\rho) g_m ^i (\rho) \partial _0 \rho \right| _0 ^ \infty - \int d \rho A _s (\rho) g_m ^i (\rho) \partial _\rho \left(\rho \partial _0 \rho \right)\nonumber\\
= \left. \rho A _s (\rho) g_m ^i (\rho) \partial _0 \rho \right| _0 ^ \infty - \int \rho d \rho A _s (\rho) g_m ^i (\rho) \frac {\partial _0 \rho} {\rho} .
\end{eqnarray}
The remaining $\rho$ integral exactly cancels the $\varphi$ integral of (\ref{pint}). The surface term contribution from the infinity is zero, simply because $A _s (\rho)$ asymptotically approaches zero towards infinity.

Putting all the above results together, the Lagrangian derived from the Lagrangian density is given only by the angular integration of the surface term evaluated at the skyrmion center
\begin{eqnarray}
- \int \rho d \rho d \varphi \partial _0 \left(\frac {n _1 s _2} {1 + n _3}\right) = \frac {2 g_m ^i (0)} {\theta_0' (0)} \int d \varphi \psi _m ^i (\varphi, t) \partial _0 \rho \nonumber\\
= \frac {2 \pi} {\theta _0 ' (0)} \left[g_1 ^i (0) \left[\dot {\rho} _c ^1 (t) b _1 ^i (t) - \dot {\rho} _c ^2 (t) a _1 ^i (t)\right] + g_ {-1} ^i (0) \left[\dot {\rho} _c ^1 (t) b _ {-1} ^i (t) + \dot {\rho} _c ^2 (t) a _ {-1} ^i (t)\right]\right],
\end{eqnarray}
which is  just the sum of the terms proportional to $v _c ^\dagger$ in the Lagrangian (\ref{lag}).

\section{Effective potentials}
\label{potential}
The explicit expressions for the effective potentials appearing in the magnon Hamiltonian are given by
\begin{eqnarray}
v _1 = \frac {h _D} {2} \left(n _3 \hat {\phi} \cdot \textbf{A} _y - \hat {r} \cdot \textbf{A} _x\right) - \textbf{A} _x \cdot \textbf{A} _y, \nonumber\\
v _0 = \frac {h _D} {2} \left(n _3 \hat {\phi} \cdot \textbf{A} _x - \hat {r} \cdot \textbf{A} _y\right) - \frac {n _1 ^2} {4} h _D ^2 - \frac {\textbf{A} _ x \cdot \textbf{A} _ x + \textbf{A} _ y \cdot \textbf{A} _ y} {2} + \frac {3 n _3 ^2 - 1} {2} + n _3 h _0,\nonumber\\
v _3 = \frac {h _D} {2} \left(n _3 \hat {\phi} \cdot \textbf{A} _x + \hat {r} \cdot \textbf{A} _y\right) - \frac {\textbf{A} _ x \cdot \textbf{A} _ x - \textbf{A} _ y \cdot \textbf{A} _ y} {2} - \frac {n _1 ^2} {2}.
\end{eqnarray}
We have defined two additional vector gauge fields $\textbf{A} _x = \hat {x} A _1 ^1 + \hat {y} A _2 ^1$, $\textbf{A} _y = \hat {x} A _1 ^2 + \hat {y} A _2 ^2$ for simplicity of notation. $\hat {r} = \hat {x} \cos \phi _0 + \hat {y} \sin \phi _0$ and $\hat {\phi} = \hat {y} \cos \phi _0 - \hat {x} \sin \phi _0$ are radial and azimuthal unit vectors in the magnetization space projected onto the 2D film plane. It can be seen that the exchange and the DM contributions to the magnon Hamiltonian is completely attributable to action of the emergent gauge field. Using cylindrical coordinates $\rho$ and $\varphi$ centered around the skyrmion center, the vector gauge fields have simple forms, $\textbf{A} _x = \hat {\varphi} n _1 / \rho$ and $\textbf{A} _y = \hat {\rho} \theta _0'$. Substitute in the expressions for the gauge fields, the effective potentials simplify to
\begin{eqnarray}
v _0 = \chi \frac {h _D} {2} \left( \frac {n _1 n _3} {\rho} - \theta _0' \right) - \left( \frac {h _D ^2} {2} + \frac {1} {\rho ^2} \right) \frac {n _1 ^2} {2} - \frac {{\theta _0'} ^2} {2} + \frac {3 n _3 ^2 - 1} {2} + n _3 h _0, \nonumber\\
v _3 = \chi \frac {h _D} {2} \left( \theta _0' + \frac {n _1 n _3} {\rho} \right) - \frac {1 + \rho ^2} {2 \rho ^2} n _1 ^2 + \frac {{\theta _0'} ^2} {2}.
\end{eqnarray}
The effective potential in multiplication with $\sigma _1$ is always zero, $v _1 = 0$, so we omitted it in the main text.

\section{Path integral using magnon coherent states}
\label{pathint}
As we do not consider the mode mixing for magnons caused by higher order terms in the magnon Lagrangian, our treatment of the path integral can be carried out for each individual magnon mode. So in the following we omit the mode indices $m$ and $i$, writing $a _m ^i$ as simple $a$, $b _m ^i$ as $b$, etc. Then in terms of the complex magnon field $c = a - i b$, the magnon Hamiltonian is $H = \omega c ^* c = \omega c ^\dagger c$, where in the last step we have promoted the coefficient $c$ to a quantum mechanical operator. This form of magnon Hamiltonian is identical to the Hamiltonian for quantum harmonic oscillators. So we can use bosonic coherent states to formulate the path integral, or propagator from time $t _i$ to $t _f$, for magnons as
\begin{eqnarray}
U (t _f, t _i) = <\psi _f |T e ^ {- i \int dt H} |\psi _i> = \int D (\bar {\psi}, \psi) e ^ {i S}\nonumber\\
= \lim _{N \rightarrow \infty} \prod _ {n = 1} ^ {N - 1} d (\bar {\psi} _n, \psi _n) e ^ {\bar {\psi} _0 \psi _0 + \sum _ {n = 0} ^ {N - 1}[(\bar {\psi} _ {n + 1} - \bar {\psi} _n) \psi _n -  i H (\bar {\psi} _ {n + 1}, \psi _n) \Delta t]},
\end{eqnarray}
where we have set $\hbar = 1$ and $ \Delta t = (t _f - t _i)/N$. $\psi _n$ denotes the complex field at time $t _n = t _i + n \Delta t$, $\bar {\psi} _n$ is the complex conjugate of $\psi _n$, and $N$ is the number of the time intervals used to discretize the path integral. The appearance of the additional $\bar {\psi} _0 \psi _0$ term in the exponent is caused by the fact that we did not consider the trace of the time evolution operator $U (t _f, t _i)$. The Lagrangian is given by $L = i \bar {\psi} \dot {\psi} - H$. Using again $a$ and $b$ to represent the real and imaginary parts of the complex field $\psi$, $\psi = a + i b$, we have $i \bar {\psi} \dot {\psi} = b \dot {a} - a \dot {b}$ by discarding total derivatives in time. Then the Lagrangian just simply reproduces the classical one, $L = b \dot {a} - a \dot {b} - \omega (a ^2 + b ^2)$.

The coupling to skyrmion motion and STT can be introduced by including an additional term in the Lagrangian $L = i \bar {\psi} \dot {\psi} - \omega \bar {\psi} \psi + i \eta (\bar {\psi} v - \bar {v} \psi)/2$, where $v = v _c ^1 + i v _c ^2$ is a complex velocity field. The coupling shown here is for the $m = 1$ modes, and for the $m = - 1$ modes it is $i \bar {\eta} (\bar {\psi} \bar {v} - v \psi)/2$. Below we will only consider explicitly the coupling to the $m = 1$ modes, and the $m = - 1$ modes can be treated exactly in the same manner. After integration the exponent becomes
\begin{eqnarray}
i S = \bar {\psi} _N e ^{- i \omega t} \psi _0 - \bar {\psi} _0 \psi _0 - \frac {\eta} {2} \left[\bar {\psi} _N e^ {- i \omega (t _f - t _i)} \int dt' e^ {i \omega t'} v (t') - \psi _0 \int dt' \bar {v} (t') e^ {- i \omega t'}\right]\nonumber \\
- \frac {\eta ^2} {4} \int dt' \bar {v} (t') e^ {- i \omega t'} \int dt'' \Theta (t' - t'') e^ {i \omega t''} v (t'').
\end{eqnarray}
The appearance of the Heaviside step function $\Theta (t)$ ensures the causality. Using the Fourier transform of the Heaviside function
\begin{equation}
\Theta (t) = \int \frac {d \omega} {2 \pi} \frac {i} {\omega + i 0 ^+} e ^{- i \omega t} = \int \frac {d \omega} {2 \pi} \left[\frac {i} {\omega} + \pi \delta (\omega)\right] e ^{- i \omega t},
\end{equation}
the double integral can be written as
\begin{eqnarray}
\int dt' dt'' \int \frac {d \omega'} {2 \pi} \bar {v} (t') \frac {i} {\omega' - \omega} e ^{- i \omega' (t' - t'')} v (t'') + \frac {1} {2} \int dt' dt'' \bar {v} (t') e ^{- i \omega (t' - t'')} v (t'')\nonumber \\
= - \frac {i} {\omega} \int dt' \bar {v} (t') v (t') + \frac {1} {2} \int dt' dt'' \bar {v} (t') e ^{- i \omega (t' - t'')} v (t'') + \cdots.
\end{eqnarray}
In the series expansion around $\omega' = 0$, we kept only the lowest order term and all higher order terms were discarded, as we are considering only the low frequency motion of skyrmions. In this case, we can assume that the Fourier component $v (\omega')$ is sharply peaked around $\omega' = 0$ and the higher order terms are negligibly small. Inclusion of higher order terms will introduce higher order derivatives on the velocity $v$.

The action then is
\begin{eqnarray}
i S = \bar {\psi} _N e ^{- i \omega t} \psi _0 - \bar {\psi} _0 \psi _0 - \frac {\eta} {2} \left[\bar {\psi} _N e^ {- i \omega (t _f - t _i)} \int dt' e^ {i \omega t'} v (t') - \psi _0 \int dt' \bar {v} (t') e^ {- i \omega t'}\right]\nonumber \\
+ i \frac {\eta ^2} {4 \omega} \int dt' \bar {v} (t') v (t') - \frac {\eta ^2} {8} \int dt'' \bar {v} (t'') \int dt' e ^{i \omega (t' - t'')} v (t').
\end{eqnarray}
The first two terms are just the free magnon action. The coupling between skyrmion motion and the remaining boundary magnon fields is described by the third term. The fourth term corresponds to a kinetic energy contribution to the action, while the last term amounts to a dissipation to the skyrmion motion.

\end{document}